# Ontology-based system to support industrial system design for aircraft assembly

**Xiaodu Hu* Rebeca Arista** Xiaochen Zheng*** Joachim Lentes* Jyri Sorvari**** Jinzhi Lu*** Fernando Ubis**** Dimitris Kiritsis*****

**Fraunhofer IAO/University Stuttgart IAT, Nobelstrasse 12, 70569 Stuttgart, Germany (Tel: +49 711 970-2337; e-mail: xiaodu.hu@iao.fraunhofer.de; joachim.lentes@iao.fraunhofer.de).*

*** Airbus SAS, 2 Rond-Point Emile Dewoitine, 31700 Blagnac, France (e-mail: rebeca.arista@airbus.com)*

**** École Polytechnique Fédérale de Lausanne (EPFL), Lausanne 1015, Switzerland (e-mail: xiaochen.zheng@epfl.ch; jinzhi.lu@epfl.ch; dimitris.kiritsis@epfl.ch)*

***** Visual Components Oy, Vänrikinkuja 2, FIN-02600 Espoo, Finland (e-mail: Jyri.Sorvari@visualcomponents.com; fernando.ubis@visualcomponents.com)*

**Abstract**: The development of an aircraft industrial system is a complex process which faces the challenge of digital discontinuity in multidisciplinary engineering due to various interfaces between different digital tools, leading to extra development time and costs. This paper proposes an ontology-based system, aiming at functionality integration and design process automation, by Models for Manufacturing methodology principles. A tool-agnostic modelling, simulation and validation platform with Discrete Event Simulation and 3D simulation is enabled and demonstrated in a real case study. An ontology layer collecting the domain knowledge enables integration of the proposed system, accelerating the design process and enhancing design quality.

## 1. INTRODUCTION

An industrial system in aerospace industry is characterized by a low-scale manufacturing rate and middle-level to high-level product customization (Arista et al. 2019a). Different scenarios of industrial systems are evaluated in the early design phase of an aircraft program, so that decisions concerning optimized architecture, lead-time, etc. can be made. Current engineering process for industrial systems design inherits model-based paradigm. Due to its limitation of developing multifaceted models (Zeigler et al. 2018) and the lack of continuity in the Product Lifecycle Management (PLM), the potential of collaboration between engineering, manufacturing and support processes are underutilized. Regarding digital continuity and interoperability, certain deficiencies in industrial system design are identified, such as different modeling languages of behavior models for different domains; manual requirement traceability; only local optimizations in industrial system design, etc.

A Model-based System Engineering (MBSE) approach is designed to resolve the aforementioned deficiencies via a unified system description using models. The International Council on Systems Engineering (INCOSE) defines MBSE as "the formalized application of modeling to support system requirements, design, analysis, verification and validation activities beginning in the conceptual design phase and continuing throughout development and later life cycle phases." (Hart 2015). Nonetheless, several gaps are still existing: (1) Knowledge management is rarely addressed by MBSE to enhance the collaborative engineering and management (Gardan und Matta 2017). Though MBSE supports system development by a structural formalisms, domain specific knowledge of a complex system represented by MBSE models cannot be captured as expected (Yang et al. 2021); (2) In aerospace industry, MBSE has been partially adopted during last few years. Although research and development have been conducted applying MBSE to the aircraft product design, the aircraft manufacturing filed is only gaining attentions recently, whereby several corresponding researches are mentioned in (Mas et al. 2018). It still lacks solutions to meet the needs of designing an industrial system for aircraft assembly; (3) Unlike the analysis and evaluation in the context of Design-to-Product, such as supporting development of an aircraft system through a Functional Mock-up Interface (FMI) based co-simulation (Hallqvist et al. 2017), Design-to-Manufacturing focuses rather on evaluation of factors in manufacturing, such as cost, lead time, etc. It requires to be addressed.

Regarding the aforementioned deficiencies and gaps, a research question is put forward: how to leverage ontology and MBSE approaches together to optimize the digital continuity and interoperability in aircraft assembly system design.

To answer the question, this paper proposes an ontology-based system to support aircraft assembly design. The main objective of this paper is to explore and validate an approach that leverages ontology as a semantic core to integrate the modeling, simulation and requirement management activities in industrial system design, in line with MBSE methodology.

The rest of this paper is organized as follows: related works to develop the system are introduced in section 2. The system architecture and functional modules are explained in section 3. Section 4 describes an implementation case. A brief conclusion and future works are summarized in section 5.

## 2. RELATED WORKS

### *2.1 Ontology-based system paradigm*

A framework for the Semantic Assembly Design Modeling with a service-oriented architecture has been proposed (Kim et al. 2006). It leverages an Assembly Design ontology to provide request service for assembly information. An EU-funded IP project PISA (Lanz et al. 2008) has defined an ontology with a knowledge base architecture to retrieve and share knowledge for simulation of manufacturing systems. Another EU-funded research project has developed an end-to-end advanced platform for manufacturing engineering and PLM (amePLM) (Lentes und Zimmermann 2017), which bases on a semantic backend and supports engineering activities with its embedded modules and the possibilities to integrate various external tools via adapter agents. Particularly in aerospace industry, a preliminary approach Models for Manufacturing (MfM) with its 3-Layers Model (3LM) – *data layer*, *ontology layer* and *service layer* – was proposed, which inherits MBSE methodology and aims for supporting the generation and management of manufacturing ontologies (Mas et al. 2018). Moreover, a Model Lifecycle Management (MLM) system supports the MfM methodology. The MLM involves three activities: ontology development, ontology lifecycle management and data structure generation for data instantiation. It seeks to design complex manufacturing systems regarding ontology layers (Arista et al. 2019b).

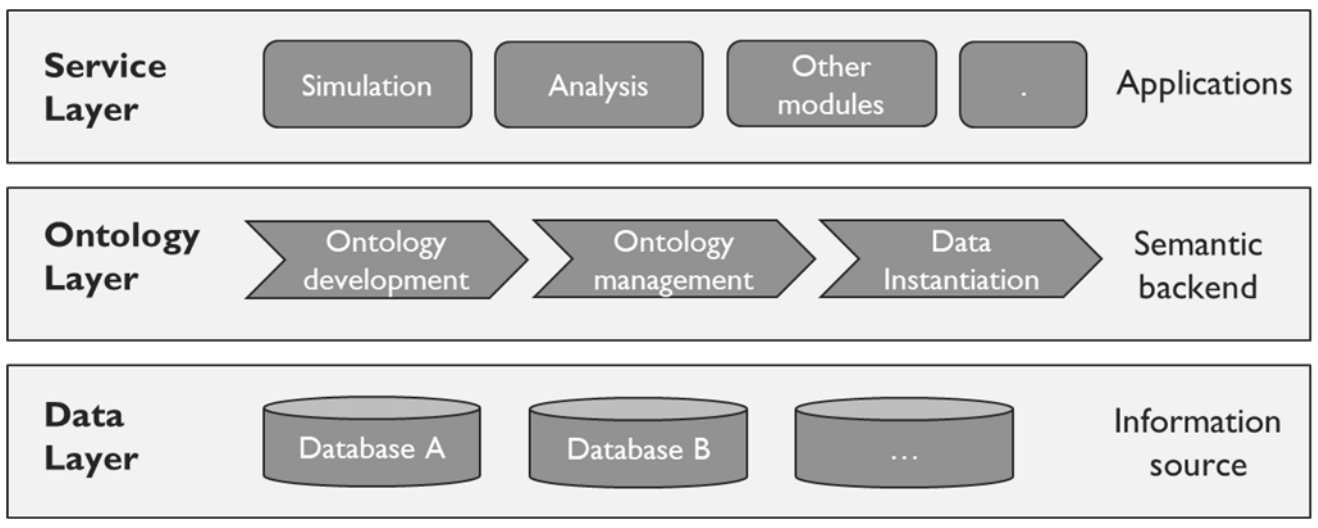

Figure 1. Paradigm of developing the ontology-based system

Based on the aforementioned research and developments among others, a general paradigm to depict an ontology-based system can be derived as knowledge acquisition, semantic processing and engineering support (Figure 1). The development work in this paper follows this paradigm.

### *2.2 Manufacturing domain ontology*

There are many ontologies for a common semantic description in manufacturing domain, such as MASON (Manufacturing's Semantics Ontology) as a preliminary upper ontology for manufacturing (Lemaignan et al. 2006), MSO (Manufacturing System Ontology) for discrete manufacturing (Negri et al. 2015), GRACE Ontology addressing quality optimization of discrete manufacturing (Foehr et al. 2013) using MPFQ-model (Material, Production Processes, Product Functions/Features, Product Quality), etc. However, ontologies in different domains are developed with different languages and tools. Recent efforts have been spent on unifying and standardizing existing domain ontologies based on certain top-level ontologies regarding manufacturing. The Industrial Ontologies Foundry (IOF) has initiated a set of open ontologies to support the manufacturing for industrial needs (Kulvatunyou et al. 2018) and provides an IOF-Core ontology. The IOF-Core refers to Basic Formal Ontology (BFO), which is a top-level ontology used in hundreds of active projects in scientific and other domains (Smith et al. 2019).

The ontology-based system in this paper relies on an application ontology that follows the IOF-Core.

### *2.3 MBSE and ontology*

Heterogeneous MBSE modeling languages and techniques lead to discrepancies among data structures and language syntaxes, which complicates information exchange among MBSE models and results in difficulties for data interoperability. To describe system characteristics and system development based on meta-models for different architecture descriptions, a unified MBSE ontology is proposed that uses meta-meta models including Graph, Object, Point, Property, Role, and Relationship (GOPPRR) (Wang et al. 2019), and in addition, the GOPPRRE with extensions (Lu et al. 2020a).

The GOPPRRE ontology is applied in this paper to mediate the data exchange between Systems Modeling Language (SysML), domain-specific knowledge, discrete event simulation (DES) models and 3D as well as ergonomic simulation models.

## 3. ONTOLOGY-BASED SYSTEM TO SUPPORT INDUSTRIAL SYSTEM DESIGN

### *3.1 System architecture with functional modules*

The proposed system is developed towards a tradespace exploration for aircraft industrial system design. Figure 2 depicts its system architecture with functional modules in 3LM.

As shown in Figure 2, the *ontology layer* is designed to import spreadsheets and structural formats of requirements and process information from *data layer*. The aim of such import is to consolidate dispersed data and information, as well as to enrich the ontology with individual knowledge of certain scenario. Besides, the process information are represented in architecture model and behaviour model of an instantiated scenario. The *service layer* provides services like simulation, design space exploration, visualize, dashboard etc. (Mas et al. 2018). In this paper, the *service layer* includes four modules: 1) the requirement management module manages the industrial requirements in requirement table; 2) the architecture definition module serves to coordinate system architecture definition in SysML models and ontology, as well as define the trades; 3) the verification module takes the needed information from *ontology layer* and pre-processes it for DES model and 3D simulation model. In the 4) visualization module, the simulation process of 3D simulation are presented and the simulation results and the insights in results are visualized in a trade-off dashboard.

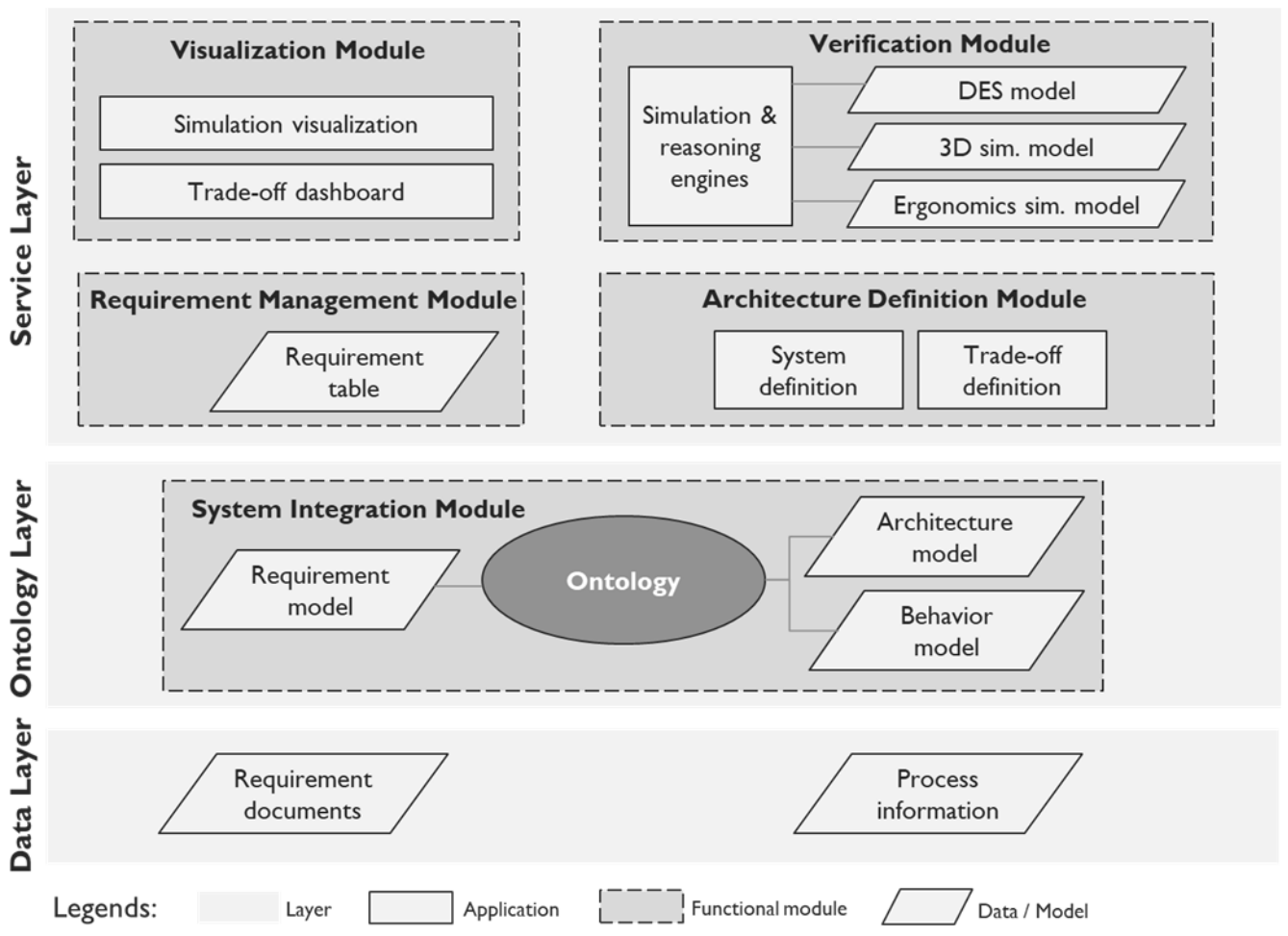

Figure 2. System architecture with functional modules in 3LM

### *3.2 System integration module*

In this module, an application ontology is developed to support ontology integration of domain-specific knowledge including system, process, and MBSE related information, such as models and requirements, as shown in Figure 4 (Zheng et al. 2020a). It uses the IOF-Core and BFO as a reference top-level ontology (Figure 5), which provides a series of key vocabularies commonly used in different industrial domains. Subclasses and individuals related to the application scenario are added accordingly to the IOF-Core ontology. They include the specific domain knowledge about the assembly process and correspond to the MBSE knowledge based on the GOPPRRE method (Lu et al. 2020a). This application ontology serves as a semantic core in the proposed ontology-based system and is used to integrate information and knowledge of the requirements, architecture and behavioral models, and process specifications.

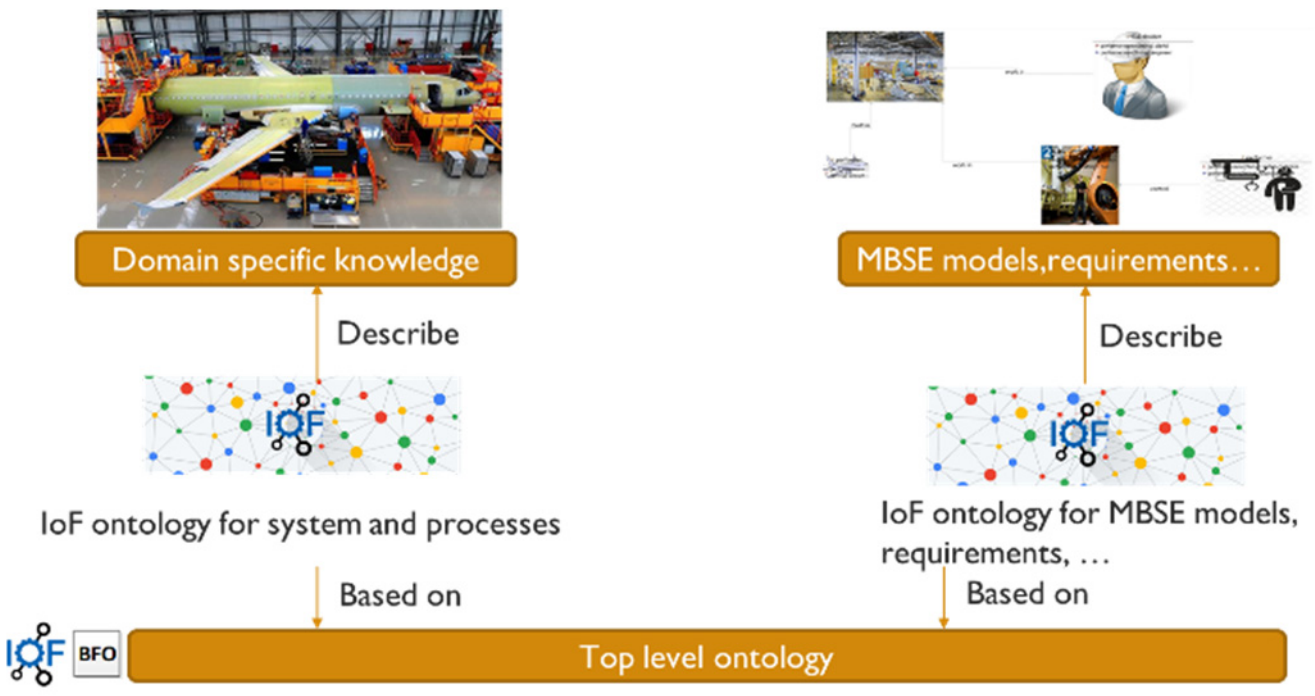

Figure 4. Ontology integration of domain specific knowledge and MBSE models

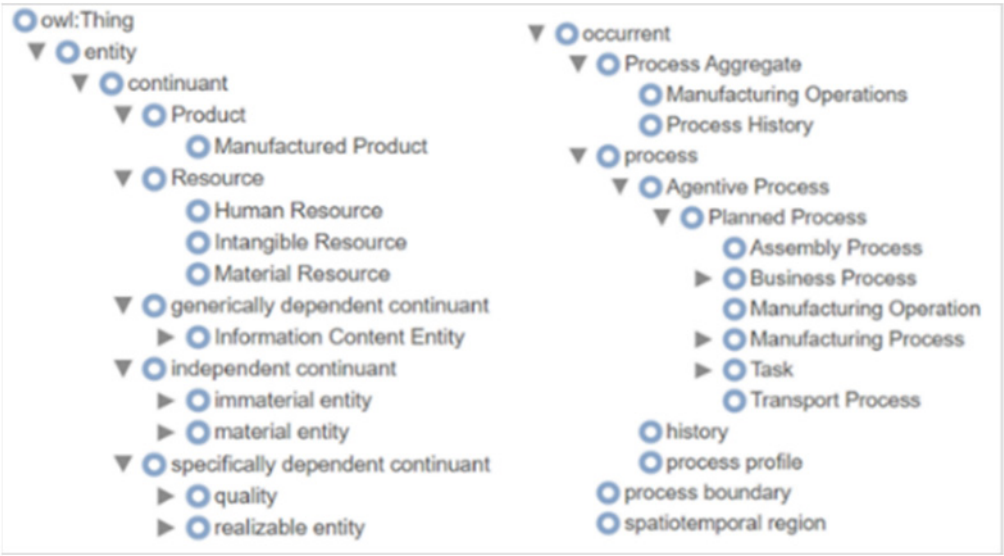

Figure 5. Main classes of the IOF-Core ontology

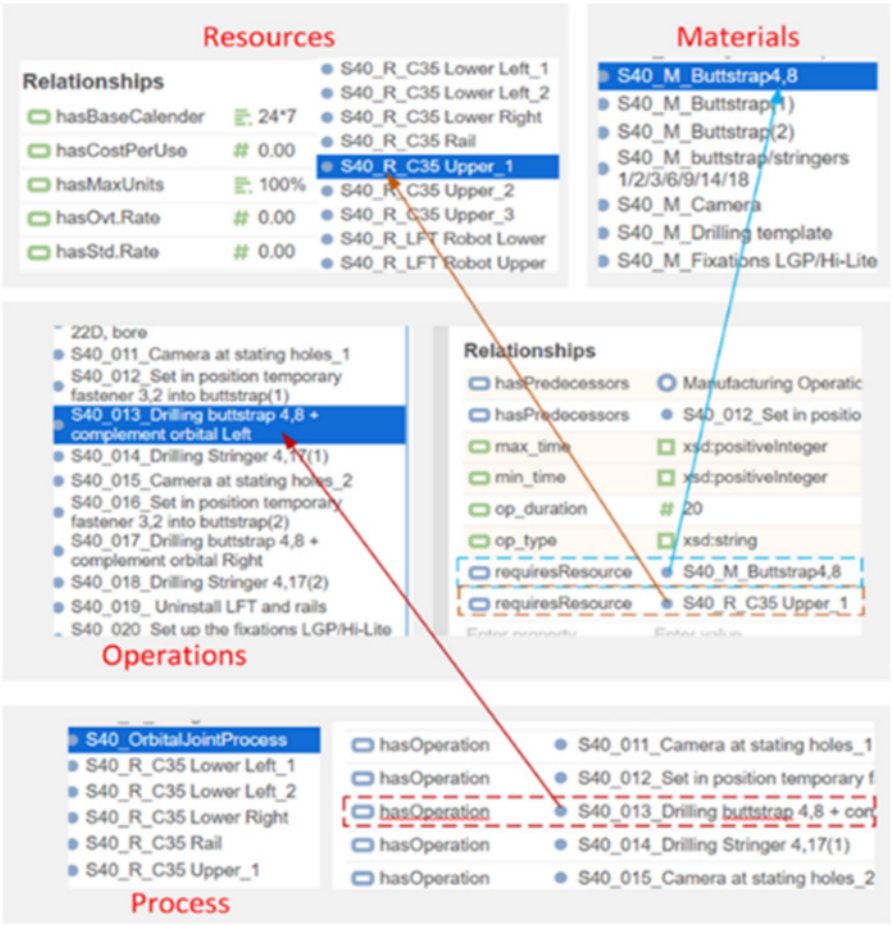

Figure 3. Generalizing individuals to create new classes

When the knowledge of aircraft assembly, e.g., an orbital joining process specification, is captured, it is necessary to analyze the relationships between frequently used classes for the knowledge instantiation. A RMPFQ (Resource, Material, Processes, Functions/Features, Quality) model (Zheng et al. 2020b) is used to organize the relevant classes and specify their relationships. Besides, each of the operations requires relevant materials and resources to execute. Figure 3 illustrates an example that an operation with the name *"S40_013_Drilling buttstrap 4.8"* has a relation *requiresMaterial* pointing to the material *"S40_M_Buttstrap4.8"*, as well as *requiresResource* *"S40_R_C35 Upper_1"*. Furthermore, each manufacturing resource and material has its own properties representing the constraints and characteristics, which are designed to support trade-offs in later steps. For instance, the *"S40_R_C35 Upper_1"* in Figure 3 has properties like *"hasBaseCalender"*, *"hasCostPerUse"* and *"hasOvt.Rate"*, representing availability, cost and efficiency of this resource.

Eventually, the application ontology is enriched by architecture information from the architecture definition module, and by industrial requirements from the requirement management module. The corresponding Web Ontology Language (OWL) file out of this module offers the input for further services at the *service layer*. The domain knowledge are managed in this semantic core.

### *3.3 Requirement management module*

The requirement management module is located at the *service layer* and provides the functionality of defining target performance requirements to support industrial engineers to evaluate the architectures and make decisions. This module aims at enabling an ontology-based definition of performance requirements based on top-level industrial requirements.

The requirement management module imports the requirements description documents from *data layer*, which can be ordinary Microsoft Office documents, Requirements Interchange Format (ReqIF) (OMG 2016) etc., through the application ontology in the *ontology layer*. Next, industrial engineers can work on requirements directly in the user interface of requirement management module, which is able to adapt the input requirements back into *ontology* and *data* *layers*, and visualize them in tables or graphs. The definition and description of those requirements, on one hand, can be exported in spreadsheets for further use; on the other hand, they can be exported as OWL and ReqIF formats, which can be integrated together with architecture definition in the application ontology in the system integration module. Moreover, the verification module can feedback the requirement management module for confirmation. It realizes the requirement traceability upon the verification states.

## 3.4 Architecture definition module

The architecture definition module is developed for defining and generating industrial system architectures at the *service layer*. The industrial design requirements should be considered while generating the architecture options and when defining the trade space to study these options. Therefore, both industrial requirements and assembly process information are to be mapped in the application ontology.

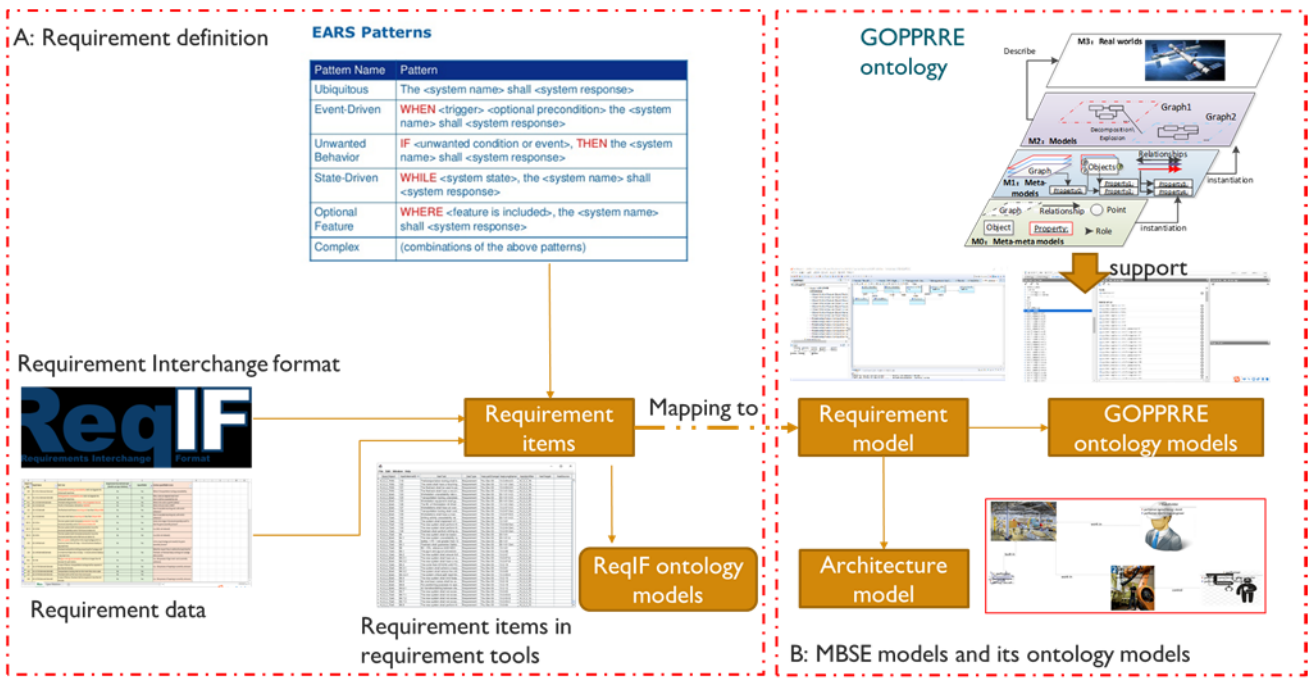

Figure 7. Mapping from requirements to architecture models

As illustrated in Figure 7, industrial requirements in ReqIF are described in requirement items, following the pattern of Easy Approach to Requirements Syntax (EARS) (Mavin et al. 2009). The EARS requirement specification provides a semi-formal requirement template in order to define the functional requirements. When the requirements are defined, they can be transformed into ReqIF ontology models, which is integrated in the system integration module.

The requirement items are mapped to requirement models, which is dependent to architecture models. In this paper, we make use of KARMA language to develop requirement and architecture models based on the GOPPRRE ontology. As shown in Figure 8, KARMA language (Lu et al. 2020b) is used to create meta-models including UPDM diagrams for formalizing the mission view and operation scenarios and SysML diagrams for formalizing the functional views, logic views and physical structure of assembling processes. KARMA language is a formal language for developing meta-models and models of MBSE, based on meta-meta models including Graph, Object, Point, Property, Relationship, Role, and extension. Moreover, in MetaGraph (Lu et al. 2020a), a plugin for GOPPRRE ontology generation is used to transform the KARMA models into OWL models based on GOPPRRE ontology. In the previous research, GOPPRR meta-meta models (Kelly et al. 1996) are investigated as one of the most powerful meta-meta modeling language which enables to support meta-model design for most of the MBSE models (Kern et al. 2011).

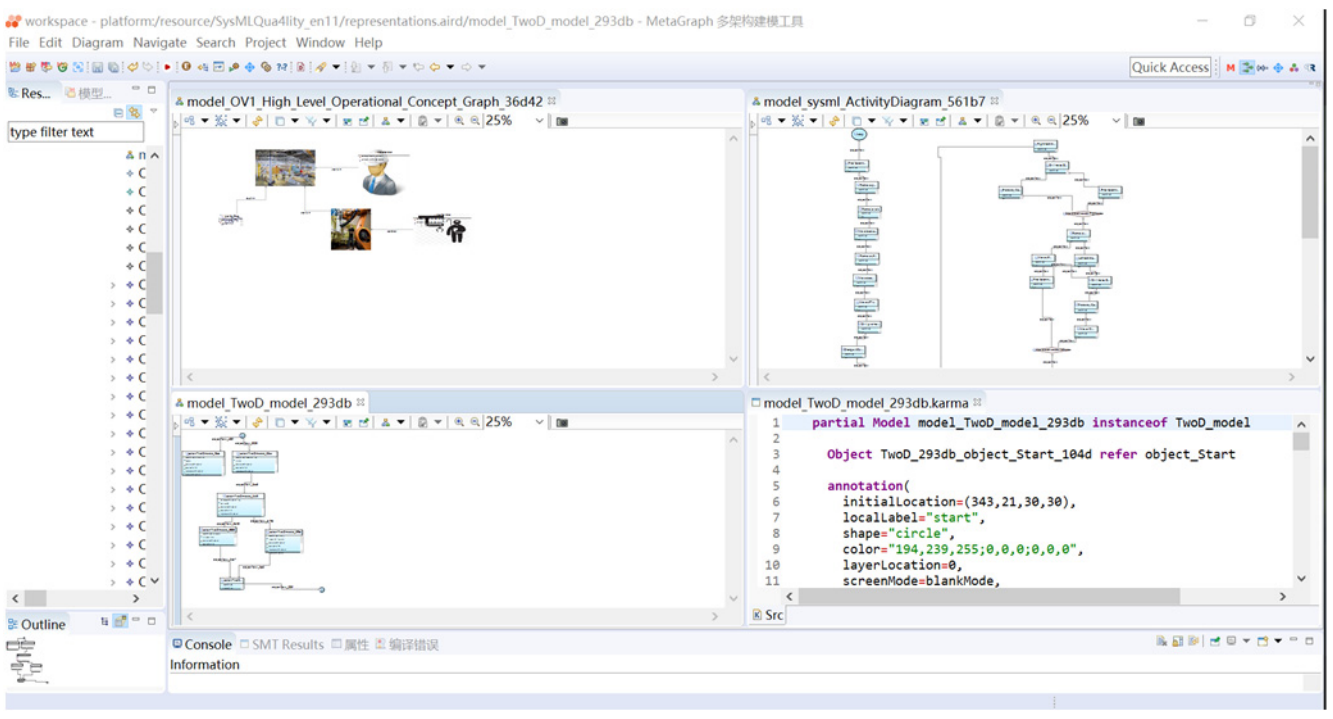

Figure 8. KARMA models for assembly process architecture design

## *3.5 Verification module*

The verification module belongs to the *service layer*. It uses the ontology as a single point of truth from *ontology layer* and supports industrial engineer by simulating industrial system designs. There are two simulations included: 1) DES and 2) 3D simulation.

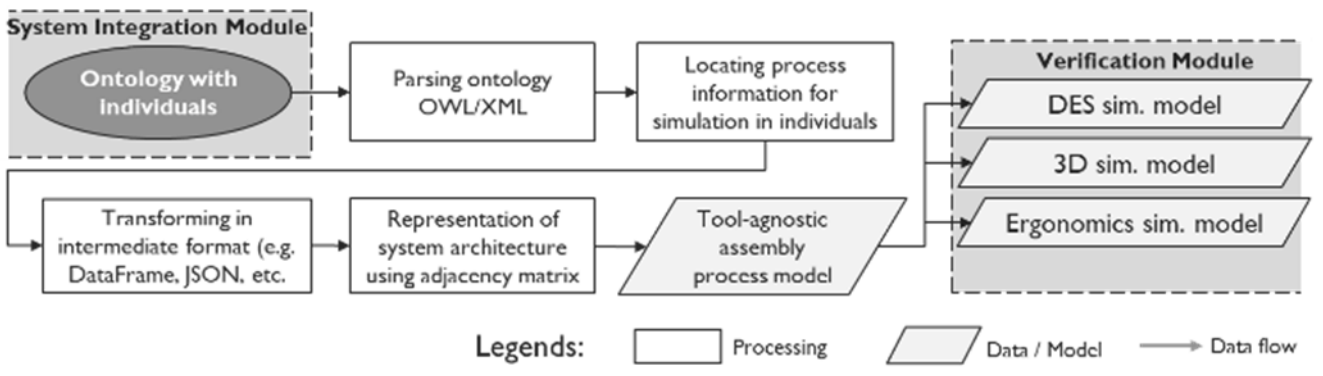

Figure 6. Procedures of utilizing ontology for verification

The procedures of utilizing ontology for verification are illustrated in Figure 6. The integrated ontology contains numerous individuals, including process and architecture information. In the first step, the ontology is parsed concerning its XML-elements and corresponding levels. Next, process information in ontology are located, and the relevant data needed for simulations are extracted. In addition, intermediate formats are used to cache the extracted process data. Later, the topology of assembly processes are represented by adjacency matrix. Finally, a tool-agnostic model is made available to be consumed by different simulation models. The final step inherits the co-modelling approach (Gomes et al. 2017) that different simulations utilizes the same model.

In Figure 6, the critical procedure is to represent the assembly process. Modeling the conceptual design of aircraft assembly bases on the decomposition of the activities, building process and resource units (Mas et al. 2013). Principally, a basic building block of information structures can be described as element and relation with their attributes (Lutters et al. 2000). Correspondingly, a data model defined in ontology for aerospace assembly lines can be enriched with many attributes and relations (Mas et al. 2019). Therefore, this paper describes the assembly process with operations and their relations, which can be represented by a vertex and an edge respectively. Thus, the topology of an assembly process can be represented by a Directed Acyclic Graph (DAG), and each assembly operation contains respective attributes. It leads to a generalized representation that also fits other scenarios in the same context.

Discrete event simulation: A DES model must contain two essential parts: 1) the duration of each operation and 2) the architecture information i.e. operation predecessor/successor.

In DAG, each operation has an in- and out-degree, which corresponds to its predecessors/successors. The DES engine recognizes the topology and simulates the lead-time, etc.

3D simulation simulates the real working environment, operation details and interactions between operators and parts. It configures the required data, e.g., operation duration, 3D model paths and coordination information. Subsequently, the routes, actions of operator and interaction between operators and parts as well as resources are simulated. In addition, by enriching the data of virtual objects, such as the weight of jig and tools, ergonomic factors can be calculated regarding operator's body posture and working time.

### *3.6 Visualization module*

The visualization module lies at the *service layer* and provides an interactive user interface to show compact information supporting industrial engineers. On one hand, it presents the simulation results; on the other hand, data analysis is performed to generate insights into the assembly system design according to engineer's needs, for instance, Gantt diagram, automation ratio, line chart for process lead-time etc.

## 4. IMPLEMENTATION

The implementation case of the ontology-based system is a trade-off between a manual assembly process and a semi-automatic assembly process using Light Flextrack (LFT) robotics. For the manual process, both drilling operations are performed by operators; for the semi-automatic process, LFT takes over the external drilling task, while the internal drilling operations are conducted manually.

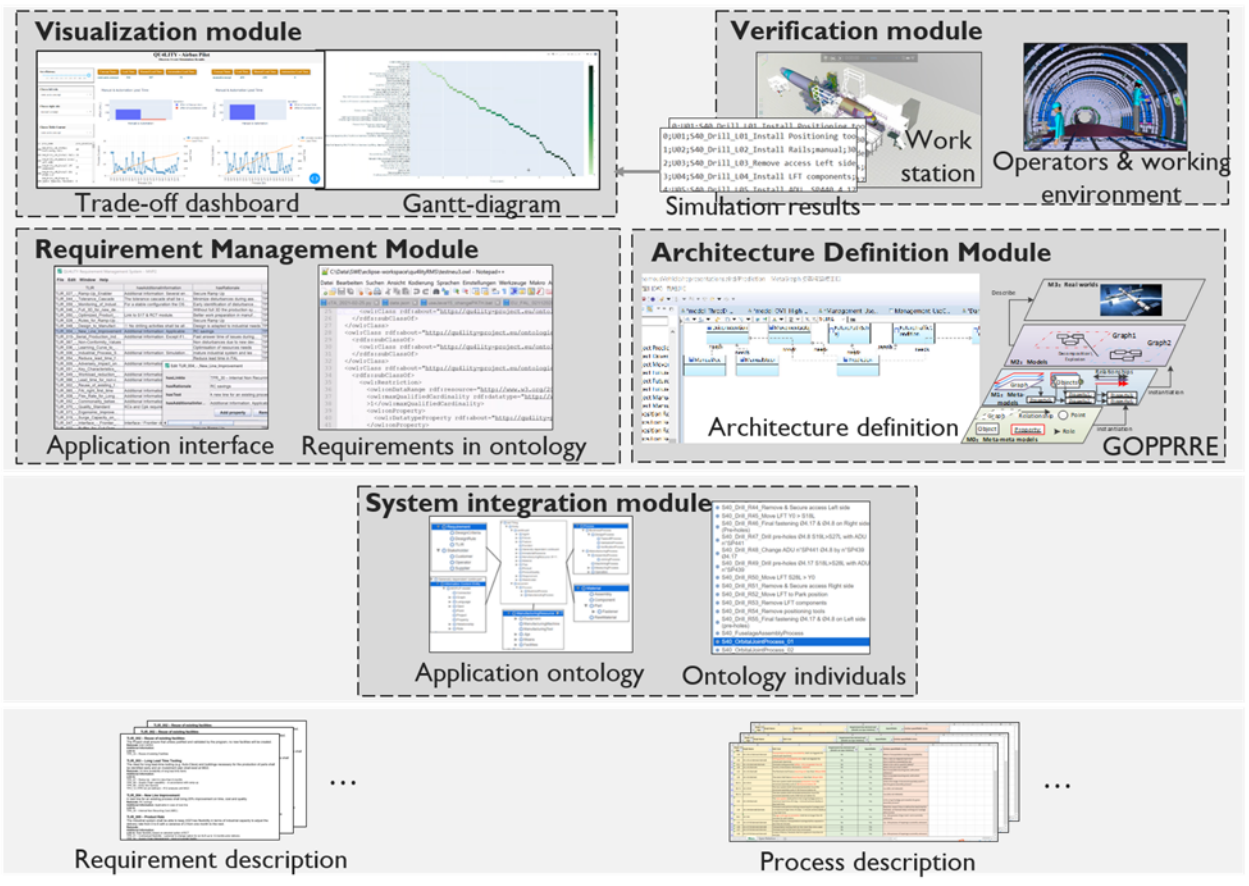

Figure 9. Simple scenario trade-off implementation

Figure 9 depicts the implementation corresponding to each functional module. At the *data layer*, the description of top-level industrial requirements and process information as required data are used to perform an industrial system design trade-off. The application ontology at the *ontology layer* are responsible for mapping the dispersed data from *data layer* and creates the corresponding OWL individuals. Subsequently, as a unified model, the ontology is fed for simulations. At the *service layer*, the OWL classes are parsed to align with different simulation models and to fit certain simulation engine. Ultimately, the metrics, e.g. lead-time, operation duration and Gantt diagram, are visualized.

## 5. CONCLUSION AND FUTURE WORKS

The implementation validates the paradigm of leveraging ontology as a semantic core with MBSE approach to integrate different models and engineering activities in industrial system design. As results, domain knowledge in the area of aircraft assembly design can be utilized and reused to generate certain ontology individuals for further usage. The proposed system achieves a requirement traceability, creates a common definition of system architecture and realizes a tool-agnostic unified modelling language to support system design. The digital continuity among engineering processes is increased, the interoperability between services and applications is enhanced, and the reusability of domain knowledge is assured.

The presented system in this paper is under development as a part of an on-going European project (QU4LITY). Future development works are addressed as follows: (1) Further enrichment of domain knowledge and establishment of knowledge formalization and persistence; (2) Design option trade space generation and exploration based on the developed ontology; (3) Further integration of functional modules and interfaces towards an automatic workflow.

## 6. ACKNOWLEDGEMENTS

The work presented in this paper is funded by the EU H2020 project QU4LITY (825030) – Digital Reality in Zero Defect Manufacturing.